\documentclass[sigconf,authorversion,nonacm]{acmart}

\usepackage{enumitem}
\usepackage{cleveref}
\usepackage[framemethod=tikz]{mdframed}
\mdfdefinestyle{textbox}{%
    roundcorner=6pt,
    backgroundcolor=teal!12,
    linewidth=1pt,
    frametitlerule=true,
    frametitlebackgroundcolor=gray!20,
    frametitlefont=\sffamily\bfseries\color{black},
    align=center,
    font=\footnotesize,
    userdefinedwidth=0.48\textwidth,
}

\usepackage[htt]{hyphenat}
\usepackage{algorithm}
\usepackage[noend]{algpseudocode}

\algrenewcommand\alglinenumber[1]{\footnotesize #1}

\AtBeginDocument{%
  \providecommand\BibTeX{{%
    \normalfont B\kern-0.5em{\scshape i\kern-0.25em b}\kern-0.8em\TeX}}}

\copyrightyear{2024}
\acmYear{2024}

\usepackage{listings}
\definecolor{darkgray}{rgb}{0.33, 0.33, 0.33}

\lstnewenvironment{Python}[1][]
  {\lstset{language=Python,
    basicstyle      = \scriptsize\ttfamily,
    keywordstyle    = \color{blue},
    keywordstyle    = [2] \color{teal}, 
    stringstyle     = \color{magenta},
    literate={ü}{{\"u}}1 {ö}{{\"o}}1 {É}{{\'E}}1 {œ}{{\oe}}1,
    commentstyle    = \color{darkgray}\ttfamily,
           morekeywords=as,
           morekeywords=with,
           #1}%
  }
  {}

\usepackage{tikz}

\newcommand*\circled[1]{\tikz[baseline=(char.base)]{\node[shape=circle,draw,inner sep=1.2pt] (char) {#1};}}

\newcommand{\header}[1]{\vspace{1mm}\noindent\textbf{#1}.}
\newcommand{\headerl}[1]{\vspace{1mm}\noindent\textit{#1}.}

\begin{document}

\title{Messy Code Makes Managing ML Pipelines Difficult? Just~Let~LLMs~Rewrite~the~Code!}

\author{Sebastian Schelter}
\affiliation{%
  \institution{BIFOLD \& TU Berlin}
  \country{}
}
\email{schelter@tu-berlin.de}

\author{Stefan Grafberger}
\affiliation{%
 \institution{BIFOLD \& TU Berlin}
 \country{}
}
\email{grafberger@tu-berlin.de}

\begin{abstract}
Machine learning (ML) applications that learn from data are increasingly used to automate impactful decisions. Unfortunately, these applications often fall short of adequately managing critical data and complying with upcoming regulations. A technical reason for the persistence of these issues is that the data pipelines in common ML libraries and cloud services lack fundamental declarative, data-centric abstractions. Recent research has shown how such abstractions enable techniques like provenance tracking and automatic inspection to help manage ML pipelines. Unfortunately, these approaches lack adoption in the real world because they require clean ML pipeline code written with declarative APIs, instead of the messy imperative Python code that data scientists typically write for data preparation.

We argue that it is unrealistic to expect data scientists to change their established development practices.  Instead, we propose to circumvent this ``code abstraction gap'' by leveraging the code generation capabilities of large language models (LLMs). Our idea is to rewrite messy data science code to a custom-tailored declarative pipeline abstraction, which we implement as a proof-of-concept in our prototype \textsc{Lester}. We detail its application for a challenging compliance management example involving ``incremental view maintenance'' of deployed ML pipelines. The code rewrites for our running example show the potential of LLMs to make messy data science code declarative, e.g., by identifying hand-coded joins in Python and turning them into joins on dataframes, or by generating declarative feature encoders from NumPy code.

\end{abstract}

\maketitle

\section{Introduction}
\label{sec:intro}

Software systems that learn from data with machine learning (ML) are increasingly used to automate impactful decisions. The risks arising from this widespread use lead to the question of how to adequately manage the data processed by these ML applications.

\header{Unsolved data management challenges in ML applications} Modern data-driven organisations run hundreds of ML pipelines to train and deploy machine learning models~\cite{xin2021production}. Unfortunately, they often fall short with respect to the management of personal and security-critical data. Google's text completion system, for example, contained credit card numbers from personal emails~\cite{carlini2019secret}, and Facebook recently could not detail the flow of personal data through its systems in a court hearing~\cite{fbdata}. At the same time, more and more regulatory requirements for ML applications are coming into effect, e.g., the right to inspect, rectify, and delete personal data from the General Data Protection Regulation in Europe, or comprehensive requirements on the traceability of results in high-risk ML applications from the upcoming European AI Act.

\header{Shortcomings in common ML libraries and cloud services} We argue that the current design of the data pipelines in such ML applications lacks the foundations to adequately address the outlined  challenges. Current ML pipeline libraries lack fundamental data-centric abstractions such as logical query plans in databases. Major cloud providers offer services based on custom pipeline abstractions~\cite{azuremlpipelines,sagemakerpipelines,vertexaipipelines} without making data a first-class citizen or modeling the semantics of individual operations. Instead, these services focus on flexibility and ease of deployment and treat the pipeline as a workflow to orchestrate with black-box operators implemented as general Python functions. Moreover, data preparation and integration are often outsourced, as a single integrated dataset is expected as input. Most of these abstractions also do neither consider the fine-grained provenance of the produced data artifacts, nor do they offer fine-grained update functionality for them. As a result, the burden of handling complex business requirements like compliance with regulations is put on the developers. 

\begin{figure*}[t!]
 \centering
 \includegraphics[width=\textwidth,page=1]{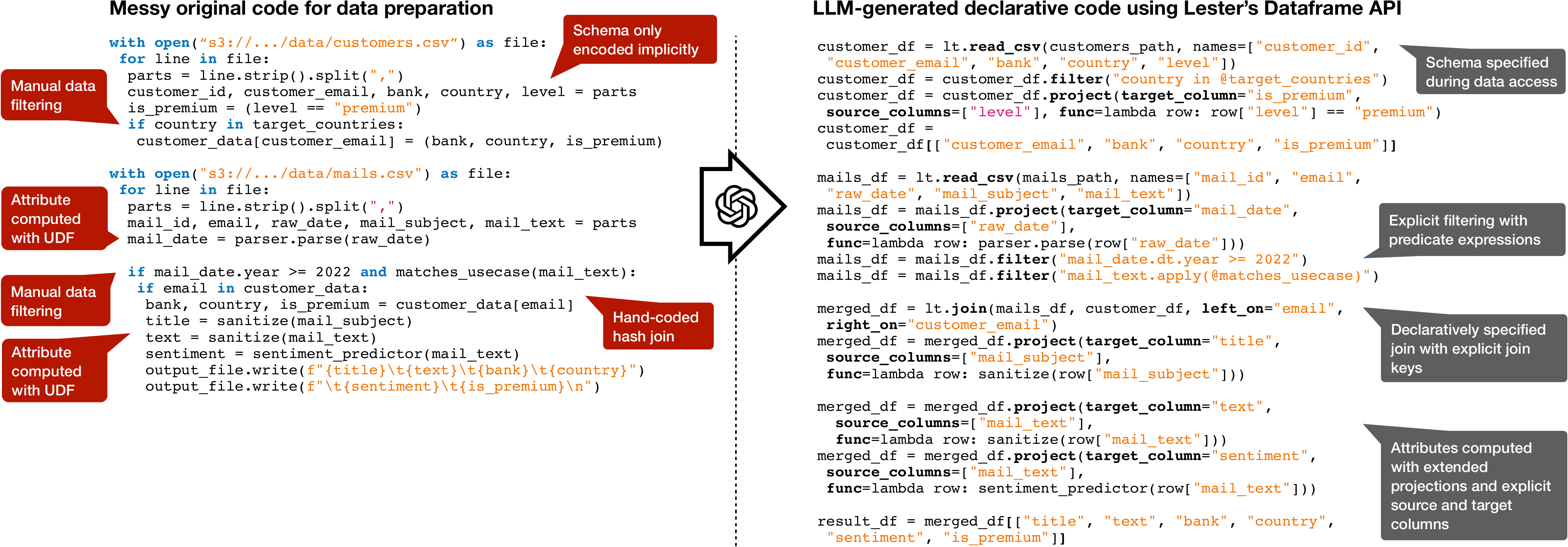}
 \caption{LLM-assisted rewrite of messy data preparation code to declarative dataframe operations in \textsc{Lester} with relational semantics and fine-grained row and column provenance tracking.}
 \label{fig:example-dataprep}
\end{figure*}

\header{Attempts to retroactively raise the level of abstraction in data science code} The data management community has recognised these challenges and shown how to enhance ML applications with provenance tracking, debugging, inspection, and automatic rewriting capabilities~\cite{namaki2020vamsa,grafberger2022data,flokas2022complaint,schelter2023proactively,grafberger2023mlwhatif}. However, these techniques are hard to integrate into real-world applications as they rely on declarative abstractions which are not present in existing data science code.

There have been various attempts to retroactively fix this. Both industry and academia have proposed novel systems~\cite{boehmsystemds,mlflowrecipes,pixeltable} for data scientists to (re)write their pipelines with. Another line of research enhances existing declaratively written ML pipelines without requiring code modifications~\cite{namaki2020vamsa,grafberger2022data,schelter2023proactively,grafberger2023mlwhatif}. Unfortunately, the real-world adoption of both of these directions is still limited. Data scientists typically focus on the ML aspects of data science and treat data preparation as grunt work for which they write messy imperative Python code to ``get the job done'' quickly. As a consequence, they lack incentives to rewrite existing code to new abstractions. Furthermore, the majority of data scientists have a math/statistics background and often lack the engineering experience to express their pipeline computations in an elaborate declarative way. 

\header{\textsc{Lester}: LLM-assisted code rewriting to a declarative pipeline abstraction} We argue that this is a fundamental problem and that it is unrealistic to expect data scientists to (re)write their ML pipeline code with declarative APIs. Instead, we propose to leverage the promising code generation capabilities of large language models (LLMs)~\cite{fernandez2023large} to bridge the outlined ``code abstraction gap''. We present a proof-of-concept in the form of our prototype \textsc{Lester}, which is based on the idea of leveraging LLMs to rewrite messy data science code to a custom-tailored declarative pipeline abstraction. We design this pipeline abstraction to encompass a variety of existing research on enhancing ML pipelines~\cite{namaki2020vamsa,grafberger2022data,flokas2022complaint,schelter2023proactively,grafberger2023mlwhatif}. 

By rewriting existing code, \textsc{Lester} can apply general functionality to address problems such as compliance with regulatory requirements. To showcase this, we implement a difficult and time-sensitive compliance task in our prototype: treating the data artifacts of a pipeline as ``materialised views'' over its inputs and conducting a low-latency update on them to delete security critical leaked input data.
\noindent In summary, we make the following contributions.
\begin{itemize}[leftmargin=*]
  \item We introduce the abstractions and implementation underlying \textsc{Lester}, discuss its code rewriting potential, and detail the resulting benefits for a challenging example (Sections~\ref{sec:example}~\&~\ref{sec:lester}).
  \item We evaluate the performance benefits of our proposed ``view maintenance'' technique and conduct a small user study to showcase that even basic tasks like computing certain metadata in ML pipelines are difficult for data scientists without system support (\Cref{sec:eval}).
  \item We provide the code of our prototype and example scenario at \textcolor{blue}{\url{https://github.com/deem-data/lester}}.
\end{itemize}

\section{Running Example}
\label{sec:example}

We introduce a running example for the scenarios that \textsc{Lester} addresses (inspired by the leakage of credit card numbers from personal emails reported in~\cite{carlini2019secret}).  Our fictitious example evolves around a financial service provider who maintains a data lake with semi-structured data of customers from several banks and their corresponding interactions, such as emails regarding questions, complaints, or service requests. Multiple ML pipelines regularly train ML models based on this data, e.g., for churn prediction, the prioritisation of service requests, text completion for chatbots, or fraud detection. 

\header{Scenario: urgent removal of security-critical personal data} Now imagine that a team of data engineers realises that credit card numbers in the email subject for customers from certain banks in Germany were not scrambled during the import of the emails into the data lake. This poses a severe financial and reputational risk for the company, so they need to act immediately to delete the leaked data.
However, the data engineers realise that several ML pipelines may have also consumed the critical data, As a result, the credit card numbers might be contained in data artifacts produced by these pipelines, might be recoverable from encoded representations using correlation attacks or might even have been memorised by the resulting ML models! The data engineers are now faced with the following challenges:

\begin{figure*}[t!]
 \centering
 \includegraphics[width=\textwidth,page=2]{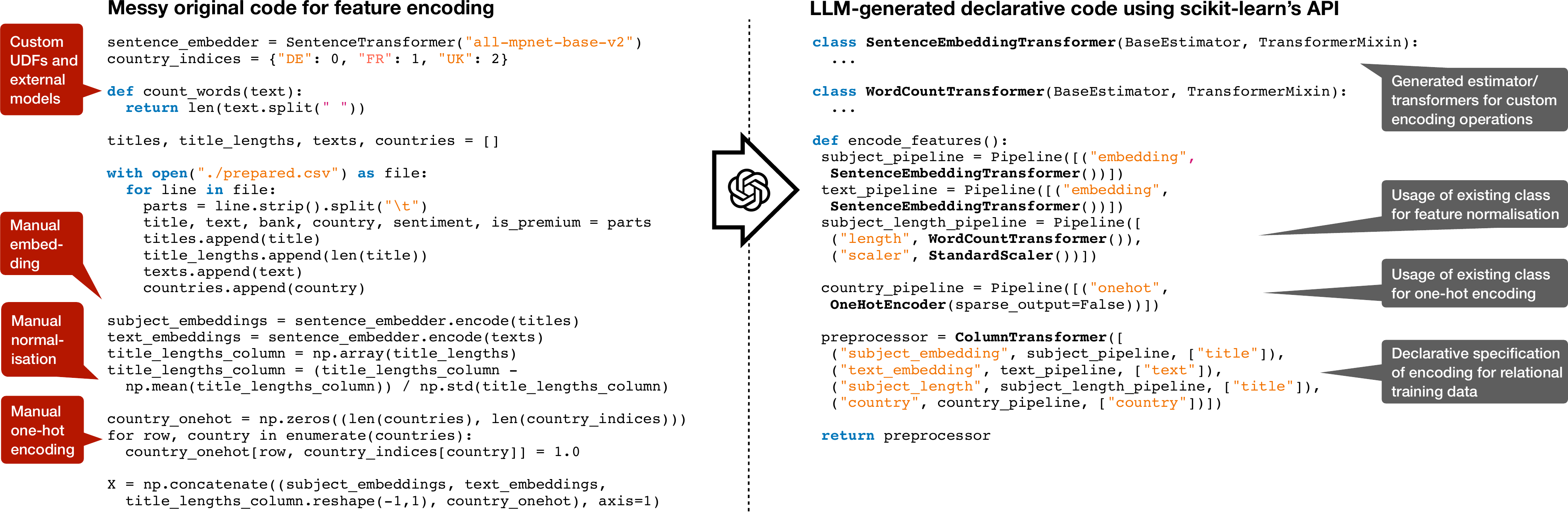}
 \caption{LLM-assisted rewrite of messy data feature encoding code to estimator/transformer operations in scikit-learn for \textsc{Lester} with fine-grained provenance tracking.}
 \label{fig:example-encode}
\end{figure*}

\begin{enumerate}[noitemsep,leftmargin=*]
  \item How can they determine which ML pipelines and models are actually affected? 
  \item If they identify the affected pipelines, how can they rectify the affected models and pipeline artifacts? They cannot simply delete all artifacts/models, as these have to be retained for compliance reasons. At the same time, it would be tedious and expensive for them to have to manually re-execute all affected pipelines from scratch to recreate the models and artifacts without the contaminated data.
\end{enumerate}

Unfortunately, the data engineers realise that the ML pipelines consist of messy Python code written by data scientists. Due to this, automated solutions for identifying affected pipelines and rectifying models and artifacts are hardly possible, and overtime hours and tedious ``detective work'' will be necessary to address the security issue!

\header{Example code} We illustrate the challenges of this scenario with messy code for an exemplary ML pipeline that identifies complaints from premium customers (based on email content). Note that the discussed pipeline code is available in our code repository at \textcolor{blue}{\url{https://github.com/deem-data/lester/blob/main/messy_original_pipeline.py}}.

\headerl{Preparing the training data} The left side of \Cref{fig:example-dataprep} (on the previous page) shows the code for preparing the training data for the ML model, based on CSV files about customers and emails from the data lake. This code has several issues, as it manually parses the CSV files, has the schema of the CSV files implicitly encoded in variable names, conducts filters and projections in plain Python code with Python UDFs and external ML models, and even conducts a hand-coded hash join to map customers to their emails. The deletion of the credit card numbers from the generated training data would require understanding column provenance, e.g., that the \texttt{title} column may contain the credit card numbers, as it is computed from the \texttt{mail\_subject} column of the input data. Furthermore, row provenance is required to identify which rows to update since the primary key (the \texttt{email} attribute) from the input data has been projected out and is not available in the training data.

\headerl{Encoding the training data as features} The code for encoding the training data as a feature matrix for model training (shown on the left side of \Cref{fig:example-encode}) is messy as well. It generates embeddings for the \texttt{title} and \texttt{text} attributes of the training data with an external model, and uses imperative NumPy code to compute the normalised word count of the \texttt{title}  as a feature and a one-hot encoding matrix of the \texttt{country} assignment.

For the deletion of the credit card numbers from the feature matrix data, a form of ``matrix column provenance'' would be required, in order to understand that the \texttt{title} column from the training data (which originates from the \texttt{mail\_subject} column with the credit card numbers) is encoded into two different features (the normalised word count and a contextual embedding). In order to update the feature matrix \texttt{X}, one also needs to be able to identify the exact dimension ranges to which these features are mapped.

\headerl{Model training} Furthermore, the pipeline trains an ML model implemented in PyTorch (whose code we omit due to lack of space, but which is available in our Github repository). 
\section{Lester}
\label{sec:lester}

In the following, we first detail the pipeline abstraction in \textsc{Lester} and showcase how to rewrite the messy code from our running example to this abstraction via custom-tailored prompts and the state-of-the-art LLM GPT-4 from OpenAI. This pipeline abstraction enables row and column provenance tracking together with certain forms of incremental view maintenance, based on which we automate the challenging compliance task described in \Cref{sec:example}.

\subsection{Computational Model for ML Pipelines} 
\label{sec:lester-model}

The backbone of \textsc{Lester} is a formal model of ML pipelines for supervised learning, which encompasses the pipeline abstractions from our previous research~\cite{grafberger2022data,schelter2023proactively,grafberger2023mlwhatif}. We model an ML pipeline as a sequence of two dataflow computations. First, the data preparation step transforms a set of relational input datasets $\mathcal{D}_1, \mathcal{D}_2, \dots, \mathcal{D}_n$ into an integrated relational dataset $\mathcal{D}_{prep}$ with a  query that applies operations from the positive relational algebra. Secondly, the feature encoding step leverages a set of featurisers $\Phi$, where $\phi_{A_i}$ denotes a sequence of estimator/transformer functions~\cite{sklearnet} for encoding the attribute $A_i$ of $\mathcal{D}_{prep}$ into matrix form. The resulting features are then concatenated to form a feature matrix $\mathbf{X}$ with a label vector $\mathbf{y}$, which denotes the training data for the pipeline's final ML model training step.

\header{Provenance-based ML pipeline management} Given this abstract pipeline representation, we can comprehensively inspect the dataflow and artifacts of the pipeline via row and column provenance tracking. We build on previous research~\cite{grafberger2022data} to efficiently track fine-grained provenance for these artifacts. In particular, we compute how-provenance via provenance polynomials~\cite{green2007provenance} during the execution of the SPJRU queries in the data preparation step and the feature encoding step. We additionally track column provenance through the projections in the relational data preparation stage, and compute a form of ``matrix column provenance'' for the feature encoding operations, where we record which column dimensions in the produced feature matrix are used by a column's feature encoders. 

\header{Implementation} We implement \textsc{Lester} as a proof-of-concept in Python. We design a dataframe API for the data preparation step with basic support for joins, selections, projections, and extended projections, which we internally execute with Pandas. We add row and column provenance tracking by maintaining a polynomial per row in a ``hidden'' column that is part of the data, and update the polynomials according to the relational operations conducted. For the feature encoding stage, we rely on the well-established estimator/transformer implementation from scikit-learn~\cite{sklearnet}, which we extend to compute ``matrix column provenance'' by determining to which dimension ranges in the feature matrix a particular input column is mapped. Based on these, we capture and store the artifacts (together with their provenance) during the initial pipeline execution.

\subsection{LLM-Assisted Code Rewrite}

In the following, we detail how to leverage OpenAI's GPT-4 model for the rewrite of the messy data science code from our running example. In general, we observe that LLMs are good at the heavy lifting (e.g., identifying handwritten joins and generating corresponding dataframe code), but that a full automation of the rewrite process is unrealistic and that data scientists have to spend additional effort on getting the code to work, e.g., by adjusting data access operations. 
Note that we focus on showcasing the potential of LLMs to turn imperative code into declarative statements, based on a set of hand-crafted prompts, and do not yet present a conversational system to automate this process. We consider it important future work to streamline and generalise the rewriting process to minimise the amount of manual corrections necessary, e.g., via an agent-based conversational approach~\cite{wu2024autogen}. 

\header{Rewriting the data preparation code to \textsc{Lester's} dataframe API} We first have the data preparation part (shown on the left side in Figure~\ref{fig:example-dataprep}) of the messy pipeline code rewritten to \textsc{Lester's} dataframe API. For that, we issue a series of prompts to the LLM that contain our transformation instructions and the messy data preparation code. The most important prompt is the first one, which asks for a rewrite to dataframe operations (and makes use of the fact that our API is similar to Pandas, for which the LLM has seen a large amount of example code during pretraining training). 

\vspace{2mm}
\begin{mdframed}[style=textbox]
The following code is written in Python with for loops and manual
data parsing. Please rewrite the code to use a dataframe library 
called lester. lester has an API similar to pandas and supports 
the following operations from pandas: \texttt{merge}, \texttt{query}, \texttt{assign}, 
\texttt{explode}, \texttt{rename}. The \texttt{assign} method in lester has two 
additional parameters: \texttt{target\_column} and \texttt{source\_columns}; 
\texttt{target\_column} refers to the new column which should be created, 
while \texttt{source\_columns} refers to the list of input columns that 
are used by the expression in \texttt{assign}. Please create a single, 
separate \texttt{assign} statement for each new column that is computed. 
Only respond with Python code. Do not iterate over dataframes. 
The code should contain a single function called 
\texttt{\_lester\_dataprep}, which returns a single dataframe called 
\texttt{result\_df} as result. This final dataframe should have the 
following columns: \texttt{<list of output columns>}\\

\noindent\texttt{<Code inserted here>}
\end{mdframed}

Note that the prompt is general and only requires filling in the desired output schema of the produced relational training data. We issue three more calls (details available in our code repository) to refine the resulting code, e.g., to make sure that references to local variables in pandas expressions are correctly annotated or that the generated code introduces function parameters for the hardcoded input paths. 

The resulting generated data preparation code is shown on the right side of \Cref{fig:example-dataprep}. We observe that the automated code rewrite successfully generates the corresponding declarative relational operations for the messy imperative data science code: $(i)$~Manual selections are now expressed with explicit calls to \texttt{filter} operations on \textsc{Lester} dataframes; $(ii)$~New columns are computed with explicit calls to \texttt{project} operations on \textsc{Lester} dataframes which contain the name of the \texttt{source\_columns} that the extended projection uses to enable column provenance tracking; $(iii)$~The handcoded hash-join is replaced with a call to \textsc{Lester's} \texttt{join} operation and explicitly specifies the join keys. Upon inspection, we find that no manual adjustments of the generated code are necessary in this example, indicating that the generation of dataframe code seems to be an easy task for LLMs.

\header{Rewriting the feature encoding code to scikit-learn's estimator/transformer API} Next, we have the LLM rewrite the feature encoding code to declarative operations using scikit-learn's estimator/transformer abstraction. We use the following prompt:

\vspace{2mm}
\begin{mdframed}[style=textbox]
The following Python reads a CSV file and manually encodes the 
data as features for a machine learning model. Please rewrite 
the code to use estimator/transformers from scikit-learn and 
the \texttt{ColumnTransformer} from scikit-learn. Only respond with 
Python code. Create a function called \texttt{encode\_features} which 
returns an unfitted \texttt{ColumnTransformer} which contains the 
feature encoding logic. The \texttt{encode\_features} function should 
be able to work on data that follows the exact schema of the 
CSV file.\\

\noindent\texttt{<Code inserted here>}
\end{mdframed}

The right side of \Cref{fig:example-encode} shows the resulting generated code for feature encoding. We again observe that the automated code rewrite successfully generates the corresponding declarative encoding operations from the messy imperative data science code: $(i)$~The manual implementations of the common normalisation and one-hot encoding operations are replaced with scikit-learns generic \texttt{StandardScaler} and \texttt{OneHotEncoder} abstractions; $(ii)$~Custom estimator/transformers are generated for the sentence embedding (\texttt{SentenceEmbeddingTransformer}) and word count feature operations (\texttt{WordCountTransformer});  $(iii)$~The complete feature encoding stage is defined declaratively with scikit-learn's \texttt{ColumnTransformer} which can directly be applied on the relational training data produced by the data preparation stage. 

For this code rewrite, two small manual adjustments are necessary to get the code to run. The sparsity parameter of the \texttt{OneHotEncoder} has to be renamed (which changed in a recent version of scikit-learn), and two additional lines of code to flatten the input lists of the custom generated transformers are needed. We assume that in future work, more sophisticated approaches (e.g., trying to execute the generated code and feeding error messages back into the LLM) and improvements in LLMs in general will help to reduce the need for manual intervention.

\begin{figure*}[t!]
 \centering
 \includegraphics[width=\textwidth]{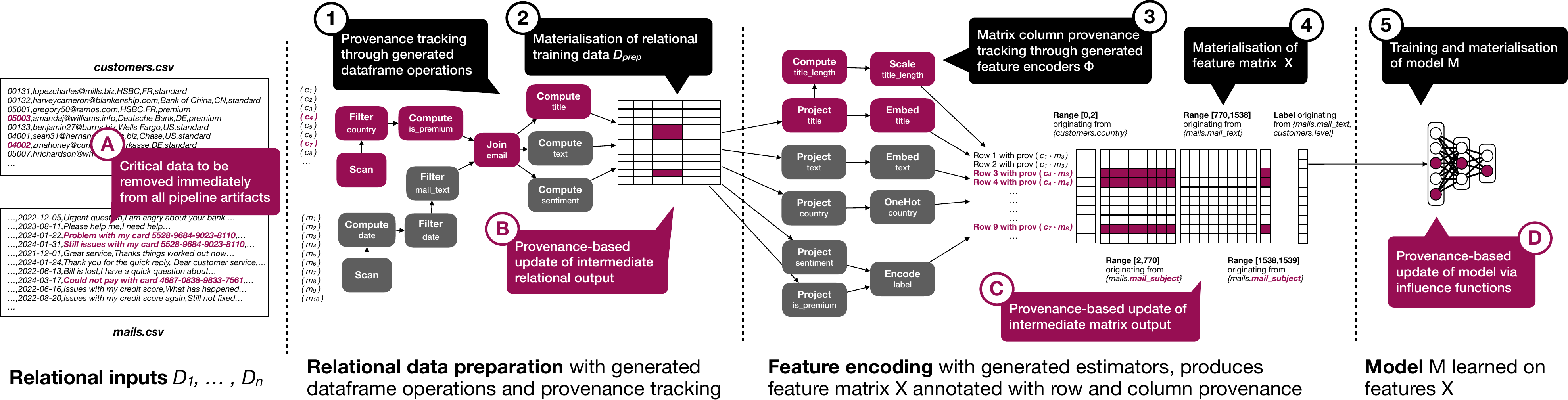}
 \caption{\protect\circled{1} -- \protect\circled{5} Initial pipeline execution with artifact capture and provenance tracking; \protect\circled{A} -- \protect\circled{D} Provenance-driven low-latency removal of security-critical data across all pipeline artifacts (intermediate training data, feature matrix, ML model).}
 \label{fig:example-unlearning}
\end{figure*}

\subsection{Incremental View Maintenance for Deployed~ML~Pipelines} 

The abstract pipeline representation at the core of \textsc{Lester} enables the automation of many common compliance tasks. As an example, we detail how to solve our outlined scenario scenario of urgently removing sensitive information leaked into a pipeline, by treating the scenario as an incremental view maintenance (IVM) problem.

\header{ML Pipelines as ``materialised views''} Our abstraction allows us to treat the produced pipeline artifacts (e.g., the relational training data, the feature matrix, and the model) as ``views'' over the relational inputs of the pipeline, which we can efficiently maintain for certain changes to the inputs. Maintaining $\mathcal{D}_{prep}$ with relational update operations and $\mathbf{X}$ with matrix operations for operations like the deletion of input records is straight-forward based on the detailed provenance information (under the reasonable assumption that small input changes do not affect the global statistics computed by estimators at fitting time). If the exact changes in the feature matrix are known, one can even do (approximate) updates of the pipeline's model based on influence functions~\cite{warneckemachine23} (which do not require retraining the model) in the common case where the model is differentiable and given in a form that allows the automated computation of gradients.

\header{IVM for our running example} The generated declarative pipeline code allows \textsc{Lester} to capture and materialise the pipeline artifacts together with their fine-grained row and column provenance during the initial execution of the pipeline at deployment time. \Cref{fig:example-unlearning} shows how \textsc{Lester} conducts a targeted low-latency update of the deployed ML pipeline for removing the security-critical data. \circled{B} For the update of the relational training data $\mathcal{D}_{prep}$, \textsc{Lester} queries the column provenance for the input column \texttt{mail\_subject} to identify that the \texttt{title} column of $\mathcal{D}_{prep}$ is computed from this, and queries the row provenance to identify the rows which originate from the affected input customers. Next, \textsc{Lester} replaces the resulting cells with NULL values to remove the leaked credit card numbers. \circled{C} The update of the feature matrix $\mathbf{X}$ works similarly. The row provenance identifies the affected rows and \textsc{Lester} queries the computed matrix column provenance to identify the dimension ranges onto which the encoded \texttt{title} column is mapped, which are subsequently overwritten with zeroes. \textsc{Lester} keeps track of the affected positions in the feature matrix together with their previous numerical values to conduct a fast first-order ``unlearning'' update of the pipeline's ML model \circled{D}, according to the general techniques proposed in~\cite{warneckemachine23}. 

\section{Evaluation}
\label{sec:eval}

We conduct a preliminary evaluation of our IVM technique and the claims on the difficulty of manually extending pipeline code.

\header{Runtime Benefits of Pipeline IVM} We evaluate the runtime benefits of our proposed IVM technique for the example scenario.

\headerl{Experimental setup} We focus on our example scenario of urgent updates of the pipeline artifacts due to the leakage of a small amount of security-critical data into the pipeline. We compare the time to re-execute the original pipeline from scratch to the time for conducting an IVM update with \textsc{Lester} on the captured artifacts. We experiment with synthetically generated customer and mail data, and ask for updates to leaked data for five customers. We experiment with a growing number (up to 100,000) of email records and customers (up to 10,000) as pipeline inputs. We execute the experiment with Python 3.9 and four AMD EPYC 7H12 2.6 GHz cores on a machine running AlmaLinux 8.6. We repeat each run seven times (except for the last run with 100,000 mails, which we only execute three times due to the long runtime) and measure the mean runtime.

\begin{figure}[h!]
  \centering
  \includegraphics[width=\columnwidth]{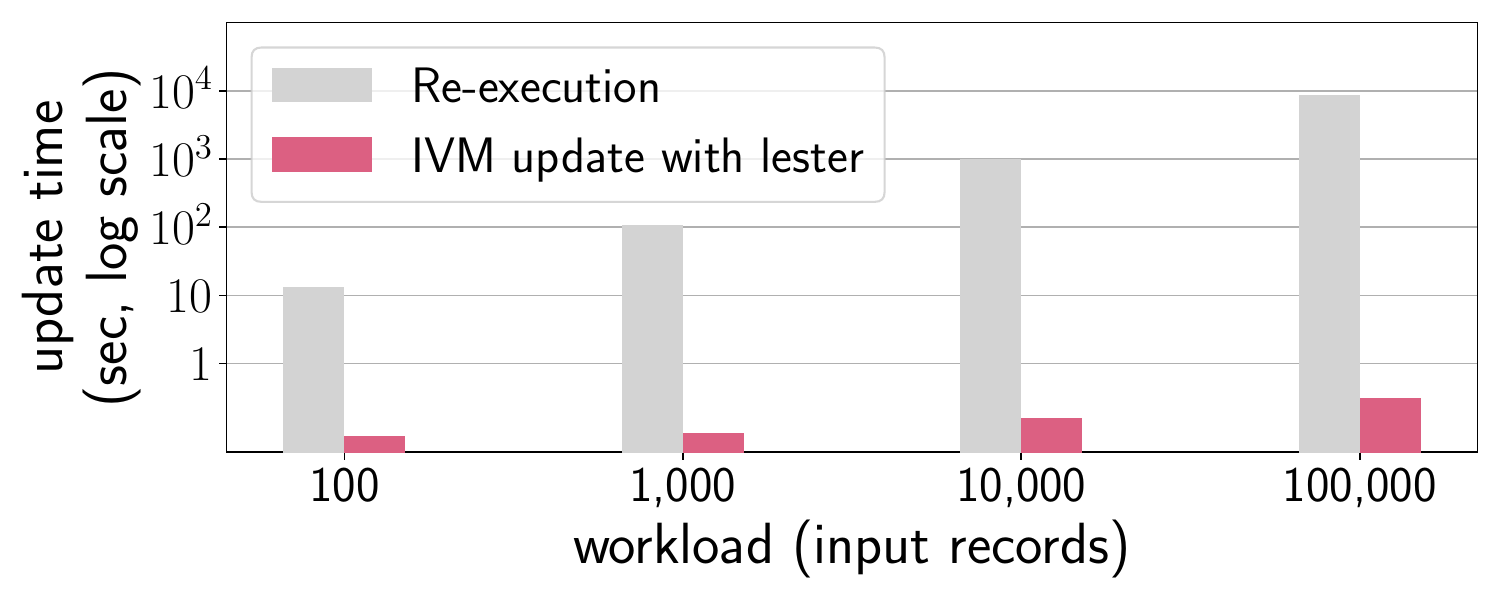}
  \caption{Time (in logarithmic scale) to re-execute a pipeline from scratch versus the time to update the pipeline artifacts with IVM. \textsc{Lester} enables sub-second updates.}
  \label{fig:ivm}
\end{figure}
\headerl{Results and discussion} We plot the resulting runtimes in seconds on a logarithmic scale in \Cref{fig:ivm}. We observe that the time to re-execute the pipeline from scratch scales linearly with the size of the input data, and that the re-execution already takes more than 140 minutes for an input size of 100,000 mails. (which is still a toy setup, compared to real-world workloads). The runtime for the IVM update of the pipeline artifacts with \textsc{Lester} is influenced by the input size, but depends on the size of the data to update and constantly ranges below a second, even for the challenging pipeline ran on 100,000 mails. These results confirm that \textsc{Lester's} design enables low-latency updates of pipeline outputs, which are orders of magnitude faster than re-execution.

\header{Exploratory User Study} Next, we conduct a small user study to showcase that even basic tasks like computing certain metadata in ML pipelines are difficult for data scientists without system support.

\headerl{Experimental setup} We design a complex ML pipeline to identify helpful product reviews in an e-commerce setting, based on a dataset of product reviews from Amazon, which includes joins, data cleaning operations, temporal splits and complex feature engineering operations. We include nine participants in our study with a mixed industry/academia background, all of which have a specialisation in ML and/or data engineering. Our participants receive a ten-minute introduction to the scenario in an online Zoom meeting, where they are given access to a Google Colab notebook with the pipeline.  

We subsequently ask them to extend the pipeline code to address the following tasks: \texttt{T1} -- Assess the group fairness (with a given fairness metric) of the pipeline for third-party reviews and non-third-party reviews; \texttt{T2} -- Track the row provenance for the products and ratings relations by computing two boolean arrays for them, which denote which records in the relations have been used to train the model.
We give the participants one hour to work on both tasks and ask them to provide us with information on the time spent per task as well as their final edited copies of the notebook. In addition, we ask them to answer eight short survey questions. We provide the materials for this study at \textcolor{blue}{\url{https://github.com/deem-data/lester/blob/main/study.md}}.

\headerl{Results and discussion} We analyse the results of our exploratory study based on the participants' notebooks and survey answers: 

\headerl{Finding 1 -- Manual extension of ML pipeline code is non-trivial} Only four out of nine participants managed to fully implement \texttt{T1} (with some minor errors), and required 35 to 56 minutes for this task. Only two of those four managed to additionally implement \texttt{T2} and spent 20 and 25 minutes on it.

\headerl{Finding 2 -- Manual code extension is error prone} We analysed the solutions of our participants in detail, and found that even the participants who stated that they successfully finished \texttt{T1} often had small mistakes in their code that lead to incorrect results. Most of these errors revolve around the implementation of the fairness metric, where we found several minor careless mistakes, e.g., copy-and-paste errors, the computation of the metric on the train instead of the test set, or hacky ways to identify group memberships that would give wrong results if other parts of the code changed. 

\headerl{Finding 3 -- Manual pipeline code extension potentially introduces high code complexity} Next, we analysed the complexity of the solutions of our participants by counting the number of lines of code (LOC) that they added to the pipeline. For \texttt{T1}, the participants added up to 37 LOC (compared to only six in our reference solution) and performed up to six additional joins on the data (compared to none in our reference solution). This indicates that even though nearly half of our participants managed to finish the implementation for this task, they introduced relatively complex code with up to seven times the number of lines required and many unnecessary joins. For \texttt{T2}, the results were closer to the reference solution with six and 11 additional LOC (compared to seven in the reference solution), and two and three additional joins (compared to two in the reference solution).

\headerl{Conclusion} In summary, our exploratory study showed that the manual extension of pipeline code is a non-trivial task (as less than half of our participants finished \texttt{T1}, and less than a quarter finished \texttt{T2}), with a high potential for errors and code complexity. This as a strong indication that data scientists need systems support and partial automation for this task.
\section{Next Steps}
\label{sec:next}

The next step for \textsc{Lester} will be a dedicated interface to guide data scientists through the LLM-assisted code rewriting process, potentially via a conversational, agent-based approach~\cite{wu2024autogen}. We also plan to design a pipeline refactoring benchmark based on code "in the wild", to get a deeper understanding of the capabilities and limitations of refactoring pipeline code with LLMs.  Furthermore, we plan to implement additional functionality for compliance management such as automated provenance-based pipeline analysis and maintenance, as well as production monitoring and logging. Furthermore, we think that the idea of using LLMs to make imperative code declarative has many applications outside of ML pipeline management, e.g., to rewrite legacy code to modern systems.



\begin{thebibliography}{20}


\ifx \showCODEN    \undefined \def \showCODEN     #1{\unskip}     \fi
\ifx \showDOI      \undefined \def \showDOI       #1{#1}\fi
\ifx \showISBNx    \undefined \def \showISBNx     #1{\unskip}     \fi
\ifx \showISBNxiii \undefined \def \showISBNxiii  #1{\unskip}     \fi
\ifx \showISSN     \undefined \def \showISSN      #1{\unskip}     \fi
\ifx \showLCCN     \undefined \def \showLCCN      #1{\unskip}     \fi
\ifx \shownote     \undefined \def \shownote      #1{#1}          \fi
\ifx \showarticletitle \undefined \def \showarticletitle #1{#1}   \fi
\ifx \showURL      \undefined \def \showURL       {\relax}        \fi
\providecommand\bibfield[2]{#2}
\providecommand\bibinfo[2]{#2}
\providecommand\natexlab[1]{#1}
\providecommand\showeprint[2][]{arXiv:#2}

\bibitem[{Amazon Web Services}(2020)]%
        {sagemakerpipelines}
\bibfield{author}{\bibinfo{person}{{Amazon Web Services}}.}
  \bibinfo{year}{2020}\natexlab{}.
\newblock \bibinfo{title}{{SageMaker Pipelines}}.
\newblock
  \bibinfo{howpublished}{\url{https://aws.amazon.com/sagemaker/pipelines/}}.
\newblock


\bibitem[{Biddle, Sam}(2022)]%
        {fbdata}
\bibfield{author}{\bibinfo{person}{{Biddle, Sam}}.}
  \bibinfo{year}{2022}\natexlab{}.
\newblock \bibinfo{title}{{Facebook Engineers: We have no idea where we keep
  all your personal data}}.
\newblock
  \bibinfo{howpublished}{\url{https://theintercept.com/2022/09/07/facebook-personal-data-no-accountability/}}.
\newblock


\bibitem[Boehm et~al\mbox{.}(2020)]%
        {boehmsystemds}
\bibfield{author}{\bibinfo{person}{Matthias Boehm} {et~al\mbox{.}}}
  \bibinfo{year}{2020}\natexlab{}.
\newblock \showarticletitle{SystemDS: A Declarative Machine Learning System for
  the End-to-End Data Science Lifecycle}.
\newblock \bibinfo{journal}{\emph{CIDR}} (\bibinfo{year}{2020}).
\newblock


\bibitem[Carlini et~al\mbox{.}(2019)]%
        {carlini2019secret}
\bibfield{author}{\bibinfo{person}{Nicholas Carlini} {et~al\mbox{.}}}
  \bibinfo{year}{2019}\natexlab{}.
\newblock \showarticletitle{The secret sharer: Evaluating and testing
  unintended memorization in neural networks}.
\newblock \bibinfo{journal}{\emph{USENIX Security}} (\bibinfo{year}{2019}).
\newblock


\bibitem[{Databricks}(2022)]%
        {mlflowrecipes}
\bibfield{author}{\bibinfo{person}{{Databricks}}.}
  \bibinfo{year}{2022}\natexlab{}.
\newblock \bibinfo{title}{{Mlflow Recipes}}.
\newblock
  \bibinfo{howpublished}{\url{https://mlflow.org/docs/latest/recipes.html}}.
\newblock


\bibitem[Fernandez et~al\mbox{.}(2023)]%
        {fernandez2023large}
\bibfield{author}{\bibinfo{person}{Raul~Castro Fernandez} {et~al\mbox{.}}}
  \bibinfo{year}{2023}\natexlab{}.
\newblock \showarticletitle{How large language models will disrupt data
  management}.
\newblock \bibinfo{journal}{\emph{VLDB}} (\bibinfo{year}{2023}).
\newblock


\bibitem[Flokas et~al\mbox{.}(2022)]%
        {flokas2022complaint}
\bibfield{author}{\bibinfo{person}{Lampros Flokas} {et~al\mbox{.}}}
  \bibinfo{year}{2022}\natexlab{}.
\newblock \showarticletitle{Complaint-driven training data debugging at
  interactive speeds}.
\newblock \bibinfo{journal}{\emph{SIGMOD}} (\bibinfo{year}{2022}).
\newblock


\bibitem[{Google}(2021)]%
        {vertexaipipelines}
\bibfield{author}{\bibinfo{person}{{Google}}.} \bibinfo{year}{2021}\natexlab{}.
\newblock \bibinfo{title}{{Vertex AI Pipelines}}.
\newblock
  \bibinfo{howpublished}{\url{https://cloud.google.com/vertex-ai/docs/pipelines}}.
\newblock


\bibitem[Grafberger et~al\mbox{.}(2022)]%
        {grafberger2022data}
\bibfield{author}{\bibinfo{person}{Stefan Grafberger} {et~al\mbox{.}}}
  \bibinfo{year}{2022}\natexlab{}.
\newblock \showarticletitle{Data distribution debugging in machine learning
  pipelines}.
\newblock \bibinfo{journal}{\emph{VLDB Journal}} \bibinfo{volume}{31},
  \bibinfo{number}{5} (\bibinfo{year}{2022}).
\newblock


\bibitem[Grafberger et~al\mbox{.}(2023)]%
        {grafberger2023mlwhatif}
\bibfield{author}{\bibinfo{person}{Stefan Grafberger} {et~al\mbox{.}}}
  \bibinfo{year}{2023}\natexlab{}.
\newblock \showarticletitle{Automating and Optimizing Data-Centric What-If
  Analyses on~Native~ Machine~Learning~Pipelines}.
\newblock \bibinfo{journal}{\emph{SIGMOD}} (\bibinfo{year}{2023}).
\newblock


\bibitem[Grafberger et~al\mbox{.}(2024)]%
        {grafberger2024red}
\bibfield{author}{\bibinfo{person}{Stefan Grafberger} {et~al\mbox{.}}}
  \bibinfo{year}{2024}\natexlab{}.
\newblock \showarticletitle{Red Onions, Soft Cheese and Data: From Food Safety
  to Data Traceability for Responsible AI.}
\newblock \bibinfo{journal}{\emph{IEEE Data Eng. Bull.}} \bibinfo{volume}{47},
  \bibinfo{number}{1} (\bibinfo{year}{2024}).
\newblock


\bibitem[Green et~al\mbox{.}(2007)]%
        {green2007provenance}
\bibfield{author}{\bibinfo{person}{Todd~J Green} {et~al\mbox{.}}}
  \bibinfo{year}{2007}\natexlab{}.
\newblock \showarticletitle{Provenance semirings}. In
  \bibinfo{booktitle}{\emph{PODS}}.
\newblock


\bibitem[{Microsoft}(2018)]%
        {azuremlpipelines}
\bibfield{author}{\bibinfo{person}{{Microsoft}}.}
  \bibinfo{year}{2018}\natexlab{}.
\newblock \bibinfo{title}{{Azure Machine Learning Pipelines}}.
\newblock
  \bibinfo{howpublished}{\url{https://learn.microsoft.com/en-us/azure/machine-learning/concept-ml-pipelines}}.
\newblock


\bibitem[Namaki et~al\mbox{.}(2020)]%
        {namaki2020vamsa}
\bibfield{author}{\bibinfo{person}{Mohammad~Hossein Namaki} {et~al\mbox{.}}}
  \bibinfo{year}{2020}\natexlab{}.
\newblock \showarticletitle{Vamsa: Automated provenance tracking in data
  science scripts}.
\newblock \bibinfo{journal}{\emph{KDD}} (\bibinfo{year}{2020}).
\newblock


\bibitem[{Pixeltable}(2024)]%
        {pixeltable}
\bibfield{author}{\bibinfo{person}{{Pixeltable}}.}
  \bibinfo{year}{2024}\natexlab{}.
\newblock \bibinfo{title}{{Pixeltable}}.
\newblock \bibinfo{howpublished}{\url{https://pixeltable.readme.io/}}.
\newblock


\bibitem[Schelter et~al\mbox{.}(2023)]%
        {schelter2023proactively}
\bibfield{author}{\bibinfo{person}{Schelter} {et~al\mbox{.}}}
  \bibinfo{year}{2023}\natexlab{}.
\newblock \showarticletitle{Proactively Screening Machine Learning Pipelines
  with ArgusEyes}.
\newblock \bibinfo{journal}{\emph{SIGMOD}} (\bibinfo{year}{2023}).
\newblock


\bibitem[{Scikit-learn}(2024)]%
        {sklearnet}
\bibfield{author}{\bibinfo{person}{{Scikit-learn}}.}
  \bibinfo{year}{2024}\natexlab{}.
\newblock \bibinfo{title}{{Pipelines and composite estimators}}.
\newblock
  \bibinfo{howpublished}{\url{https://scikit-learn.org/stable/modules/compose.html}}.
\newblock


\bibitem[Warnecke et~al\mbox{.}(2023)]%
        {warneckemachine23}
\bibfield{author}{\bibinfo{person}{Alexander Warnecke} {et~al\mbox{.}}}
  \bibinfo{year}{2023}\natexlab{}.
\newblock \showarticletitle{Machine Unlearning of Features and Labels}.
\newblock \bibinfo{journal}{\emph{NDSS}} (\bibinfo{year}{2023}).
\newblock


\bibitem[Wu et~al\mbox{.}(2024)]%
        {wu2024autogen}
\bibfield{author}{\bibinfo{person}{Qingyun Wu} {et~al\mbox{.}}}
  \bibinfo{year}{2024}\natexlab{}.
\newblock \showarticletitle{AutoGen: Enabling Next-Gen LLM Applications via
  Multi-Agent Conversation}.
\newblock \bibinfo{journal}{\emph{ICLR Workshop on Large Language Model
  Agents}} (\bibinfo{year}{2024}).
\newblock


\bibitem[Xin et~al\mbox{.}(2021)]%
        {xin2021production}
\bibfield{author}{\bibinfo{person}{Doris Xin} {et~al\mbox{.}}}
  \bibinfo{year}{2021}\natexlab{}.
\newblock \showarticletitle{Production machine learning pipelines: Empirical
  analysis and optimization opportunities}.
\newblock \bibinfo{journal}{\emph{SIGMOD}} (\bibinfo{year}{2021}).
\newblock


\end{thebibliography}


\begin{thebibliography}{20}


\ifx \showCODEN    \undefined \def \showCODEN     #1{\unskip}     \fi
\ifx \showDOI      \undefined \def \showDOI       #1{#1}\fi
\ifx \showISBNx    \undefined \def \showISBNx     #1{\unskip}     \fi
\ifx \showISBNxiii \undefined \def \showISBNxiii  #1{\unskip}     \fi
\ifx \showISSN     \undefined \def \showISSN      #1{\unskip}     \fi
\ifx \showLCCN     \undefined \def \showLCCN      #1{\unskip}     \fi
\ifx \shownote     \undefined \def \shownote      #1{#1}          \fi
\ifx \showarticletitle \undefined \def \showarticletitle #1{#1}   \fi
\ifx \showURL      \undefined \def \showURL       {\relax}        \fi
\providecommand\bibfield[2]{#2}
\providecommand\bibinfo[2]{#2}
\providecommand\natexlab[1]{#1}
\providecommand\showeprint[2][]{arXiv:#2}

\bibitem[{Amazon Web Services}(2020)]%
        {sagemakerpipelines}
\bibfield{author}{\bibinfo{person}{{Amazon Web Services}}.}
  \bibinfo{year}{2020}\natexlab{}.
\newblock \bibinfo{title}{{SageMaker Pipelines}}.
\newblock
  \bibinfo{howpublished}{\url{https://aws.amazon.com/sagemaker/pipelines/}}.
\newblock


\bibitem[{Biddle, Sam}(2022)]%
        {fbdata}
\bibfield{author}{\bibinfo{person}{{Biddle}}.}
  \bibinfo{year}{2022}\natexlab{}.
\newblock \bibinfo{title}{{Facebook Engineers: We have no idea where we keep
  all your personal data}}.
\newblock
  \bibinfo{howpublished}{\url{https://theintercept.com/2022/09/07/facebook-personal-data-no-accountability/}}.
\newblock


\bibitem[Boehm et~al\mbox{.}(2020)]%
        {boehmsystemds}
\bibfield{author}{\bibinfo{person}{Boehm} {et~al\mbox{.}}}
\newblock \showarticletitle{SystemDS: A Declarative Machine Learning System for
  the End-to-End Data Science Lifecycle}.
\newblock \bibinfo{journal}{\emph{CIDR}} (\bibinfo{year}{2020}).
\newblock


\bibitem[Carlini et~al\mbox{.}(2019)]%
        {carlini2019secret}
\bibfield{author}{\bibinfo{person}{Carlini} {et~al\mbox{.}}}
\newblock \showarticletitle{The secret sharer: Evaluating and testing
  unintended memorization in neural networks}.
\newblock \bibinfo{journal}{\emph{USENIX Security}} (\bibinfo{year}{2019}).
\newblock


\bibitem[{Databricks}(2022)]%
        {mlflowrecipes}
\bibfield{author}{\bibinfo{person}{{Databricks}}.}
  \bibinfo{year}{2022}\natexlab{}.
\newblock \bibinfo{title}{{Mlflow Recipes}}.
\newblock
  \bibinfo{howpublished}{\url{https://mlflow.org/docs/latest/recipes.html}}.
\newblock


\bibitem[Fernandez et~al\mbox{.}(2023)]%
        {fernandez2023large}
\bibfield{author}{\bibinfo{person}{Fernandez} {et~al\mbox{.}}}
\newblock \showarticletitle{How LLMs will disrupt data
  management}.
\newblock \bibinfo{journal}{\emph{VLDB}} (\bibinfo{year}{2023}).
\newblock


\bibitem[Flokas et~al\mbox{.}(2022)]%
        {flokas2022complaint}
\bibfield{author}{\bibinfo{person}{Flokas} {et~al\mbox{.}}}
\newblock \showarticletitle{Complaint-driven training data debugging at
  interactive speeds}.
\newblock \bibinfo{journal}{\emph{SIGMOD}} (\bibinfo{year}{2022}).
\newblock


\bibitem[{Google}(2021)]%
        {vertexaipipelines}
\bibfield{author}{\bibinfo{person}{{Google}}.} \bibinfo{year}{2021}\natexlab{}.
\newblock \bibinfo{title}{{Vertex AI Pipelines}}.
\newblock
  \bibinfo{howpublished}{\url{https://cloud.google.com/vertex-ai/docs/pipelines}}.
\newblock


\bibitem[Grafberger et~al\mbox{.}(2022)]%
        {grafberger2022data}
\bibfield{author}{\bibinfo{person}{Grafberger} {et~al\mbox{.}}}
\newblock \showarticletitle{Data distribution debugging in ML
  pipelines}.
\newblock \bibinfo{journal}{\emph{VLDBJ}} (\bibinfo{year}{2022}).
\newblock


\bibitem[Grafberger et~al\mbox{.}(2023)]%
        {grafberger2023mlwhatif}
\bibfield{author}{\bibinfo{person}{Grafberger} {et~al\mbox{.}}}
\newblock \showarticletitle{Automating and Optimizing Data-Centric What-If
  Analyses on~Native~ Machine~Learning~Pipelines}.
\newblock \bibinfo{journal}{\emph{SIGMOD}} (\bibinfo{year}{2023}).
\newblock




\bibitem[Green et~al\mbox{.}(2007)]%
        {green2007provenance}
\bibfield{author}{\bibinfo{person}{Green} {et~al\mbox{.}}}
\newblock \showarticletitle{Provenance semirings}.
  \bibinfo{booktitle}{\emph{PODS}} (\bibinfo{year}{2007}).
\newblock


\bibitem[{Microsoft}(2018)]%
        {azuremlpipelines}
\bibfield{author}{\bibinfo{person}{{Microsoft}}.}
  \bibinfo{year}{2018}\natexlab{}.
\newblock \bibinfo{title}{{Azure Machine Learning Pipelines}}.
\newblock
  \bibinfo{howpublished}{\url{https://learn.microsoft.com/en-us/azure/machine-learning/concept-ml-pipelines}}.
\newblock


\bibitem[Namaki et~al\mbox{.}(2020)]%
        {namaki2020vamsa}
\bibfield{author}{\bibinfo{person}{Namaki} {et~al\mbox{.}}}
\newblock \showarticletitle{Vamsa: Automated provenance tracking in data
  science scripts}.
\newblock \bibinfo{journal}{\emph{KDD}} (\bibinfo{year}{2020}).
\newblock


\bibitem[{Pixeltable}(2024)]%
        {pixeltable}
\bibfield{author}{\bibinfo{person}{{Pixeltable}}.}
  \bibinfo{year}{2024}\natexlab{}.
\newblock \bibinfo{title}{{Pixeltable}}.
\newblock \bibinfo{howpublished}{\url{https://pixeltable.readme.io/}}.
\newblock


\bibitem[Schelter et~al\mbox{.}(2023)]%
        {schelter2023proactively}
\bibfield{author}{\bibinfo{person}{Schelter} {et~al\mbox{.}}}
\newblock \showarticletitle{Screening ML Pipelines
  with ArgusEyes}.
\newblock \bibinfo{journal}{\emph{SIGMOD}} (\bibinfo{year}{2023}).
\newblock


\bibitem[{Scikit-learn}(2024)]%
        {sklearnet}
\bibfield{author}{\bibinfo{person}{{Scikit-learn}}.}
  \bibinfo{year}{2024}\natexlab{}.
\newblock \bibinfo{title}{{Pipelines and composite estimators}}.
\newblock
  \bibinfo{howpublished}{\url{https://scikit-learn.org/stable/modules/compose.html}}.
\newblock


\bibitem[Warnecke et~al\mbox{.}(2023)]%
        {warneckemachine23}
\bibfield{author}{\bibinfo{person}{Warnecke} {et~al\mbox{.}}}
\newblock \showarticletitle{Machine Unlearning of Features and Labels}.
\newblock \bibinfo{journal}{\emph{NDSS}} (\bibinfo{year}{2023}).
\newblock


\bibitem[Wu et~al\mbox{.}(2024)]%
        {wu2024autogen}
\bibfield{author}{\bibinfo{person}{Wu} {et~al\mbox{.}}}
\newblock \showarticletitle{AutoGen: Enabling Next-Gen LLM Applications via
  Multi-Agent Conversation}.
\newblock \bibinfo{journal}{\emph{ICLR Workshop on Large Language Model
  Agents}} (\bibinfo{year}{2024}).
\newblock


\bibitem[Xin et~al\mbox{.}(2021)]%
        {xin2021production}
\bibfield{author}{\bibinfo{person}{Xin} {et~al\mbox{.}}}
\newblock \showarticletitle{Production machine learning pipelines: Empirical
  analysis and optimization opportunities}.
\newblock \bibinfo{journal}{\emph{SIGMOD}} (\bibinfo{year}{2021}).
\newblock


\end{thebibliography}

\end{document}